\documentclass{IEEEtran}
\usepackage{graphicx}
\usepackage{amsmath}
\usepackage{multicol}
\usepackage{multirow}
\usepackage{stfloats}
\usepackage{booktabs}
\usepackage{mathrsfs}
\usepackage{amsmath}
\usepackage{enumerate}
\usepackage{color}
\usepackage{url}
\usepackage{verbatim}
\usepackage{amssymb}
\usepackage{dcolumn}
\usepackage{algorithm}
\usepackage{algpseudocode}

\usepackage{array}
\usepackage{underscore}
\usepackage{url}

\usepackage[draft]{hyperref}
\hypersetup{hidelinks}
\usepackage{flushend}

\usepackage[square, comma, sort&compress, numbers]{natbib}
\usepackage{graphicx}
\usepackage{epstopdf}
\usepackage{caption}
\usepackage{diagbox}

\usepackage{caption}

\usepackage{subfigure}
\usepackage{color}

\usepackage[outerbars,color]{changebar}
\ifx\pdfoutput\undefined
\else\ifnum\pdfoutput>0
  \usepackage{pdfcolmk}
\fi\fi
\cbcolor{red}
\usepackage{fancyhdr}

\makeatletter
\newcommand{\Rmnum}[1]{\expandafter\@slowromancap\romannumeral #1@}
\makeatother

\usepackage{array}  \newcommand{\PreserveBackslash}[1]{\let\temp=\\#1\let\\=\temp}  \newcolumntype{C}[1]{>{\PreserveBackslash\centering}p{#1}}  \newcolumntype{R}[1]{>{\PreserveBackslash\raggedleft}p{#1}}  \newcolumntype{L}[1]{>{\PreserveBackslash\raggedright}p{#1}}
\ifCLASSINFOpdf
\else
\fi
\begin{document}
\bibliographystyle{IEEEtran}
\bibliographystyle{unsrt}
\title{Mobility-Enhanced Simultaneous Lightwave Information and Power Transfer}

\long\def\symbolfootnote[#1]#2{\begingroup%
\def\thefootnote{\fnsymbol{footnote}}\footnote[#1]{#2}\endgroup}
\renewcommand{\thefootnote}{\fnsymbol{footnote}}
\author{Mingqing~Liu,
Mingliang~Xiong,
and Qingwen~Liu*,~\IEEEmembership{\normalsize Senior Member,~IEEE\normalsize}

%
\thanks{
\emph{Mingqing Liu, Mingliang Xiong, and Qingwen Liu are with College of Electronics and Information Engineering, Tongji University, Shanghai, People's Republic of China (e-mail: clare@tongji.edu.cn, xiongml@tongji.edu.cn, qliu@tongji.edu.cn).}}
}


\maketitle

\begin{abstract}
Simultaneous lightwave information and power transfer (SLIPT) has been regarded as a promising technology to deal with the ever-growing energy consumption and data-rate demands in the Internet of Things (IoT). We propose a resonant beam based SLIPT system (RB-SLIPT), which deals with the conflict of high deliverable power and mobile receiver positioning with the existing SLIPT schemes. At first, we establish a mobile transmission channel model and depict the energy distribution in the channel. Then, we present a practical design and evaluate the energy/data transfer performance within the moving range of the RB-SLIPT. Numerical evaluation demonstrates that the RB-SLIPT can deliver $5$W
charging power and enable $1.5$Gbit/s achievable data rate with
the moving range of $20$-degree field of view (FOV) over $3$m
distance. Thus, RB-SLIPT can simultaneously provide high-power energy and high-rate data transfer, and mobile receiver positioning capability.
\end{abstract}


\IEEEpeerreviewmaketitle

\section{Introduction}
\label{sec:Introduction}
\begin{figure}[t]
    \centering
    \includegraphics[width=3.2in]{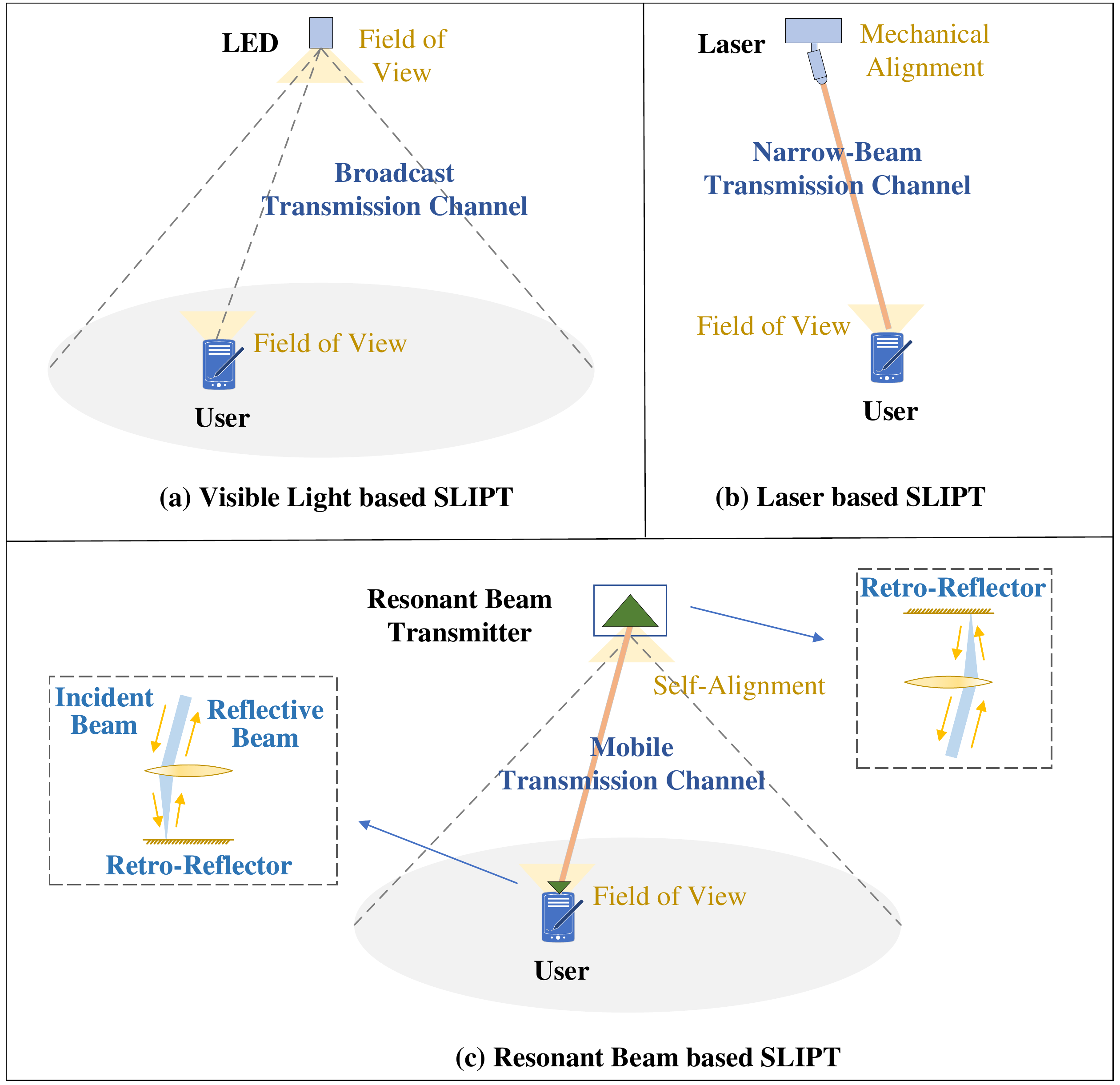}
    \caption{Comparison of the resonant beam based SLIPT with the existing SLIPT schemes.}
    \label{f:scenario}
\end{figure}

Simultaneous wireless information and power transfer has been regarded as one of the enabling technologies in 6G networks~\cite{6G}. Due to the spectrum crisis of Radio Frequency (RF), the SLIPT adopting visible light or lasers as carriers has become a promising altenative/complementary technology~\cite{SLIPT1,SLIPT2}. However, as in Fig.~\ref{f:scenario}, the visible-light-based SLIPT faces challenges of low deliverable charging power~\cite{VLC1,VLC2}, and the laser-based SLIPT is difficult to position mobile receivers~\cite{LaserSLIPT1,LaserSLIPT2}. Thus, we propose an RB-SLIPT scheme, which can deal with the conflict between high deliverable power and mobile receiver positioning. The RB-SLIPT system is capable of simultaneously providing multi-Watt charging power and high-rate data transfer with self-alignment capability over multi-meters distances.

The RB-SLIPT system inherits the characteristics of the resonant beam system (RBS). RBS is essentially an open-cavity laser resonator, where a transmitter consists of  high-reflective mirror and a gain medium, a receiver consists of an output coupling mirror and a Photovoltaic (PV) cell are spatially separated to form a resonator ~\cite{Liu2016Charging,RBComXiong,WWang2018}. The resonant beam generated within the resonator acts as the carrier to transfer energy and information over the air. Thus, the narrow resonant beam can carry high power similar to the laser, and will cease immediately once a foreign object invasions into the resonator, which brings the features of high transmission power over long range with the premise of human safety~\cite{RBCSafety}.

Moreover, two retro-reflectors are adopted at both ends of the RBS. As in Fig.~\ref{f:scenario}, retro-reflector can reflect the incident beam back parallel to the original direction regardless of the incident direction. RB-SLIPT with retro-reflectors such as corner-cube reflector (CCR)~\cite{arnold1979method,Zhou:82} and cat's eye~\cite{cateye1} features with self-alignment. Therefore, receivers moving within the coverage of RB-SLIPT can be supplied with wireless energy and data through the mobile transmission channel. The mobility feature of RBS has been discussed in~\cite{Liu2016Charging}. However, the model of the mobile transmission channel with self-alignment has not been established and the moving effects on energy/data transfer performance have not been analyzed. Thus, we analyze the energy distribution of the mobile transmission channel and propose a practical design of RB-SLIPT.



In this paper, we at first present the RB-SLIPT architecture. After revealing the self-alignment mechanism using transfer matrix method, we adopt the resonator mode analysis and laser output power calculation to depict the energy distribution in the mobile transmission channel. Then, we propose a practical RB-SLIPT design with a receiver adopting both PV and avalanche photodiodes (APD) for energy and data harvesting respectively,  and evaluate the system performance. The RB-SLIPT system can simultaneously provide $5$W wireless charging power and $1.5$Gbit/s achievable data rate within the moving range of $3$m distance and $20$-degree FOV.

\begin{figure}[t]
    \centering
    \includegraphics[width=3.55in]{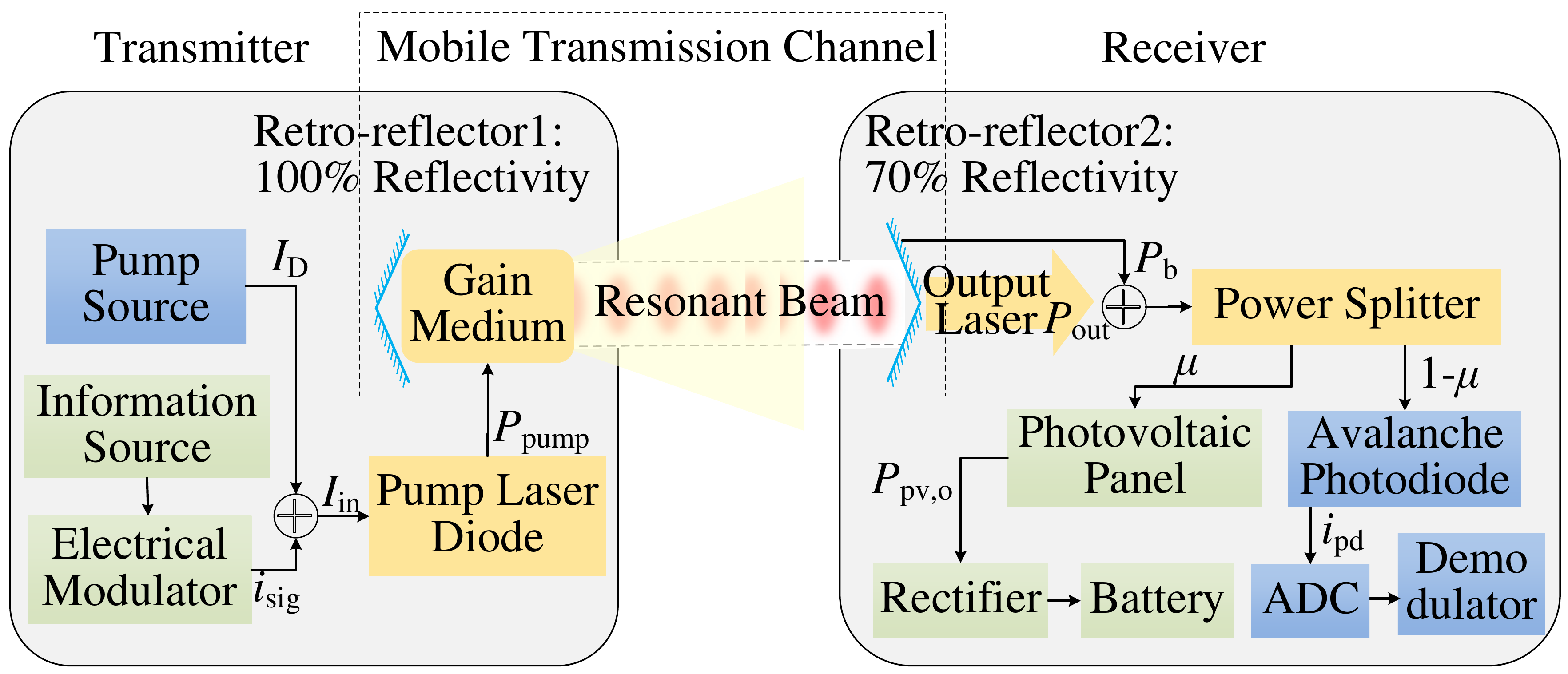}
    \caption{Transceiver design of the resonant beam based simultaneous lightwave information and power transfer system.}
    \label{f:overview}
\end{figure}

The contributions of this manuscript are:

\begin{itemize}
\item[1)] We establish a mobile transmission channel model in RB-SLIPT. It can reveal self-alignment mechanism and depict the energy distribution in the channel with a single RB-SLIPT transmitter.

\item[2)] We propose a practical design of simultaneous energy and information transfer using both PV and APD. It can quantitatively evaluate the performance of data/enrgy transfer with the impacts of moving factors.
\end{itemize}

The remainder of this paper is as follows. In Section II, we proposed the RB-SLIPT design and described its architecture and principle. In Section III, we built a mobile transmission channel model to prove the self-alignment and depict the energy distribution in the channel. Afterwards, we established an analytical model of mobile energy and information transfer in RB-SLIPT in Section IV. In Section V, we demonstrated the channel factor, charging power, and achievable data rate of the proposed system through numerical analysis. Finally, we made a conclusion in Section VI.

\section{System Overview}
\label{sec:SysOverview}

Figure~\ref{f:overview} depicts the transceiver design of the RB-SLIPT system. The proposed system consists of the spatially separated transmitter and receiver, between which the mobile transmission channel is formed.

The transmitter contains a pump source, an information source, an electrical modulator, a pump laser diode (LD), a gain medium, and a retro-relector1 with 100\% reflectivity. In the transmitter, an alternating-current (AC) current $i_{\text{sig}}$ is generated from the information source and the electrical modulator, which is biased by a direct-current (DC) current $I_{\text{D}}$ generated by the pump source. The current $I_{\text{D}}$ plus $i_{\text{sig}}$ as $I_{\text{in}}$ drives the pump LD together to generate the pump laser with power of $P_{\text{pump}}$. Then, the pump laser provides energy for the gain medium to stimulate light radiation carrying data and energy, similar to the RF amplifier. A retro-relector1 with 100\% reflectivity is adopted to reflect the light from the receiver back towards the incident direction.

The receiver contains a retro-reflector2 with 70\% reflectivity, a PV panel, an APD, and corresponding energy/data processing units. The output laser with power of $P_{\text{out}}$ is output from the retro-reflector2 with 70\% reflectivity. Then together with the background irradiance $P_{\text{b}}$ caused by the ambient light, the output power enters a power splitter to be split into two streams with a specific ratio. One stream is sent to PV and converted to the charging current with $P_{\text{pv,o}}$. After being rectified, the power is ready for charging the battery. One stream is sent to APD to be converted to the signal current $i_{\text{sig,o}}$. The signal current carries information enters the demodulator for demodulation after passing the analog-to-digital converter (ADC).

The mobile transmission channel is formed by two  retro-reflectors and a gain medium. Similar to the traditional laser, the gain medium stimulates the light radiation, which can be reflected back and forth inside the resonant cavity and pass through the gain medium multiple times to be amplified. If the light power gain can compensate for the light power loss, the narrow laser can be stably output with high-power. On the other hand, the mobile transmission channel is actually an open-cavity, and the intra-cavity laser named as resonant beam is used to transfer energy/data over the air. Thus, with a resonant cavity that allows light to be retro-reflected within it and enough input power, the resonant beam can be self-established between the transceiver. The two retro-reflectors in the mobile transmission channel form a mobile resonant cavity (MRC). Due to the retro-reflective characteristics of the two retro-reflectors, the resonant beam can be established within the MRC even as the two reflectors are not strictly facing each other, which guarantees the self-misalignment.

\section{Mobile Transmission Channel Model}
\label{sec:Mobility}
Mobile transmission channel of RB-SLIPT relies on the MRC to realize mobility. In this section, we investigate the MRC with two cat's eye at both ends as a paradigm. We at first prove the self-alignment characteristic of the MRC, after which we derive that the MRC with two deflected reflectors is equivalent to a Fabry-Perot (FP) resonator. Then, we adopt laser mode analysis method and laser output principle to obtain the power output and depict the energy distribution in the mobile transmission channel.
\subsection{Self-Alignment Mechanism of the Channel}
We adopt the transfer matrix (i.e., ABCD matrix) method to define a round-trip transfer process of a ray inside the resonant cavity. Each component (including the free space) that a ray will pass through is described as a matrix with four elements; for example, a lens with focal distance $f$ is described as $ \mathbf{M}=\left[\begin{array}{ll}
1 & 0 \\
-1/f & 1
\end{array}\right] $, and a ray represented by vector $\mathbf{r}_1 $ after passing it is transferred to $\mathbf{r}_2 $ as
 \begin{equation}
    \label{e:raytransfer}
    \mathbf{r}_{2} = \mathbf{M}\mathbf{r}_{{1}} ,
\end{equation}
where $\mathbf{r}_1 = \left[\begin{array}{ll}
r_1 \\
r_1'
\end{array}\right]$. $r_1$ is the distance between the ray and optic axis, and $r_1'$ is the slope of the ray about optic axis.

Then, a round-trip transfer matrix of a ray inside a resonant cavity is the multiplication of the matrices that a ray will pass through in a round trip, i.e., $ \mathbf{M}_{tot} =\mathbf{M}_{1}\mathbf{M}_{2} ...\mathbf{M}_{N}$, where $N$ is twice the number of components in the resonant cavity. Suppose a ray $\mathbf{r}_0 $ can realize self-reproduction after a round-trip transfer, i.e.,
 \begin{equation}
    \label{e:onaxis-axis}
    \mathbf{M}_{tot} \mathbf{r}_{{0}}=\mathbf{r}_{{0}} ,
\end{equation}
then $\mathbf{r}_0 $ can be regarded as a vector representing the axis of the cavity. Rays parallel to the axis can be reflected back and forth inside the resonant cavity.

\begin{figure}[t]
    \centering
    \includegraphics[width=3.3in]{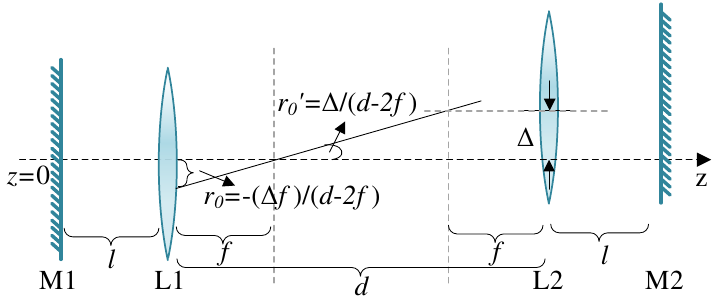}
    \caption{Self-alignment of the mobile resonant cavity.}
    \label{f:ABCDEF}
\end{figure}

As in Fig.~\ref{f:ABCDEF}, the MRC of RB-SLIPT consists of two cat's eye, where a cat's eye contains a mirror (M1/M2) and a lens (L1/L2) parallel to each other. The focal length of the lens is $f$, and the distance between lens and mirror is $l$. The cat's eye is ideal to retro-reflect incident beam as $l=f$. $z$-axis which passes through the centers of L1 and M1 is defined as the origin optic axis of the MRC. Once the right-hand cat's eye consisting of L2 and M2 is off the $z$-axis, the MRC is defined as an off-axis system.

In an off-axis system, the misalignment vector of off-axis component is defined as
\begin{equation}
    \label{e:misalignmentvector}
    \mathbf{\Delta} = \left[\begin{array}{ll}
\Delta \\
\Delta'
\end{array}\right],
\end{equation}
where $\Delta$ is the distance between element axis of off-axis component and origin optic axis, and $\Delta'$ is the slope of the element axis about the origin optic axis. Then, the relationship of $\mathbf{r}_1 $ and $\mathbf{r}_2 $ before and after passing through an off-axis component can be depicted as~\cite{siegmanlaser}
\begin{equation}
\label{e:raytransfer2}
\mathbf{ r}_{2}= \mathbf{ M} \mathbf{ r}_{1}+\mathbf{E },
\end{equation}
where
\begin{equation}
\mathbf{E} \equiv\left[\begin{array}{l}
{E} \\
{F}
\end{array}\right]=\left[\mathbf{M}_{\Delta}-\mathbf{M}\right] \mathbf{\Delta},
\end{equation}
where $\mathbf{M}_{\Delta}$ represents the optical length deference matrix of the off-axis component. Thus, the transfer process of \eqref{e:raytransfer2} can be depicted as an ABCDEF matrix~\cite{ABCDEF}:
\begin{equation}
\left[\begin{array}{l}
r_{2} \\
r_{2}^{\prime} \\
1
\end{array}\right]=\left[\begin{array}{lll}
A & B & E \\
C & D & F \\
0 & 0 & 1
\end{array}\right] \times\left[\begin{array}{c}
r_{1} \\
r_{1}^{\prime} \\
1
\end{array}\right] .
\end{equation}

As the case in Fig.~\ref{f:ABCDEF},
suppose a ray $\mathbf{r}_0 $ starts from the front face of L1, $\mathbf{r}_0 $ can be self-reproduced after a round-trip transfer process inside the MRC as
\begin{equation}
    \label{e:offaxis-axis}
\mathbf{M}_{tot} \mathbf{r}_{{0}}+\mathbf{E}_{tot}=\mathbf{r}_{{0}}, \end{equation}
then $\mathbf{r}_0 $ represents the new optic axis of MRC after the moving of the right-hand cat's eye, where~\cite{ABCDEF}
\begin{equation}
\label{e:axissolve}
r_{0} \equiv \frac{(1-D) E+B F}{2-A-D} \quad , \quad r_{0}^{\prime} \equiv \frac{C E+(1-A) F}{2-A-D} .
\end{equation}

To figure out $\mathbf{M}_{tot}$ and $\mathbf{E}_{tot}$, we expand \eqref{e:offaxis-axis} as follows:
\begin{equation}
\begin{aligned}
\mathbf{r}_{1} &=\mathbf{M}_{f s} \mathbf{r}_{0} \\
\mathbf{r}_{2}=& \mathbf{M}_{c a t} \mathbf{r}_{1}+\mathbf{E}_{M} \\
\mathbf{r}_{3} &=\mathbf{M}_{f s} \mathbf{r}_{2} \\
\mathbf{r}_{0} &=\mathbf{M}_{c a t} \mathbf{r}_{3}
\end{aligned},
\end{equation}

where

\begin{equation}
\mathbf{M}_{fs}=\left[\begin{array}{ll}
1 & d \\
0 & 1
\end{array}\right] ,
\end{equation}

\begin{equation}
\begin{aligned}
\mathbf{M}_{cat}=&\left[\begin{array}{ll}
1 & 0 \\
-\frac{1}{f} & 1
\end{array}\right]\left[\begin{array}{ll}
1 & l \\
0 & 1
\end{array}\right]\\ &\left[\begin{array}{cc}
1 & l \\
0 & 1
\end{array}\right]\left[\begin{array}{cc}
1 & 0 \\
-\frac{1}{f}  & 1
\end{array}\right] =\left[\begin{array}{ll}
-1 & 2f \\
0 & -1
\end{array}\right]
\end{aligned},
\end{equation}

\begin{equation}
\begin{aligned}
\mathbf{E}_M  =\left( \left[\begin{array}{ll}
1 & 2l \\
0& 1
\end{array}\right]  -\mathbf{M}_{cat} \right) \left[\begin{array}{ll}
\Delta \\
0\end{array}\right] =  \left[\begin{array}{ll}
-\frac{2\Delta l}{f} \\
\frac{2\Delta (1-f)}{f^2} \end{array}\right]
\end{aligned}.
\end{equation}

Then, we have~\cite{siegmanlaser}
\begin{equation}
\begin{aligned}
\mathbf{M}_{t o t} &=\mathbf{M}_{c a t} \mathbf{M}_{f s} \mathbf{M}_{c a t} \mathbf{M}_{f s} \\
\mathbf{E}_{t o t} &=\mathbf{M}_{c a t} \mathbf{M}_{f s}\left(\mathbf{E}_{M}\right)
\end{aligned}.
\end{equation}
According to \eqref{e:axissolve}, we can obtain
\begin{equation}
r_{0} = \frac{-(\Delta f)}{d-2f} \quad \text { and } \quad r_{0}^{\prime} = \frac{\Delta}{d-2f} .
\end{equation}
Evidently, $\mathbf{r}_0$ is coincident with the connection of the focuses of two cat's eye, which represents the new optic axis of the MRC with one off-axis cat's eye. Rays parallel to the new optic axis can be retro-reflected inside the MRC rather than splitting over it. Thus, MRC with two cat's eye is capable of realizing self-alignment with new optic axis even if two cat's eye are not exactly facing each other due to the movement.

As in Fig.~\ref{f:Equivalency}, the line a2 passing though the focuses of both cat's eyes and the gain medium represents a ray reflected back and forth collinearly inside the MRC. Due to the off-axis of the right-hand cat's eye, the angle between a2 and the normal vector of the cat's eye front face is $\theta$. a1 is another ray which is generated by the gain medium and parallel to a2. Then, a1 is retro-reflected by the right-hand cat's eye through points A, D, and C back towards the left-hand cat's eye along a3. Thus, there exists a resonant beam generated by and within the size of the gain medium between the two cat's eye, and the beam is symmetric about the new optic axis.

Moreover, according to the Fermat principle, the optical length of all rays inside the MRC are the same. Therefore, as the free space transmission distance is significantly larger than $f$, the MRC can be regarded as an FP resonator where the two end mirrors are located at the two focuses, and perpendicular to the new optic axis of the MRC.
\subsection{Energy Distribution in the Channel}
\begin{figure}[t]
    \centering
    \includegraphics[width=3.3in]{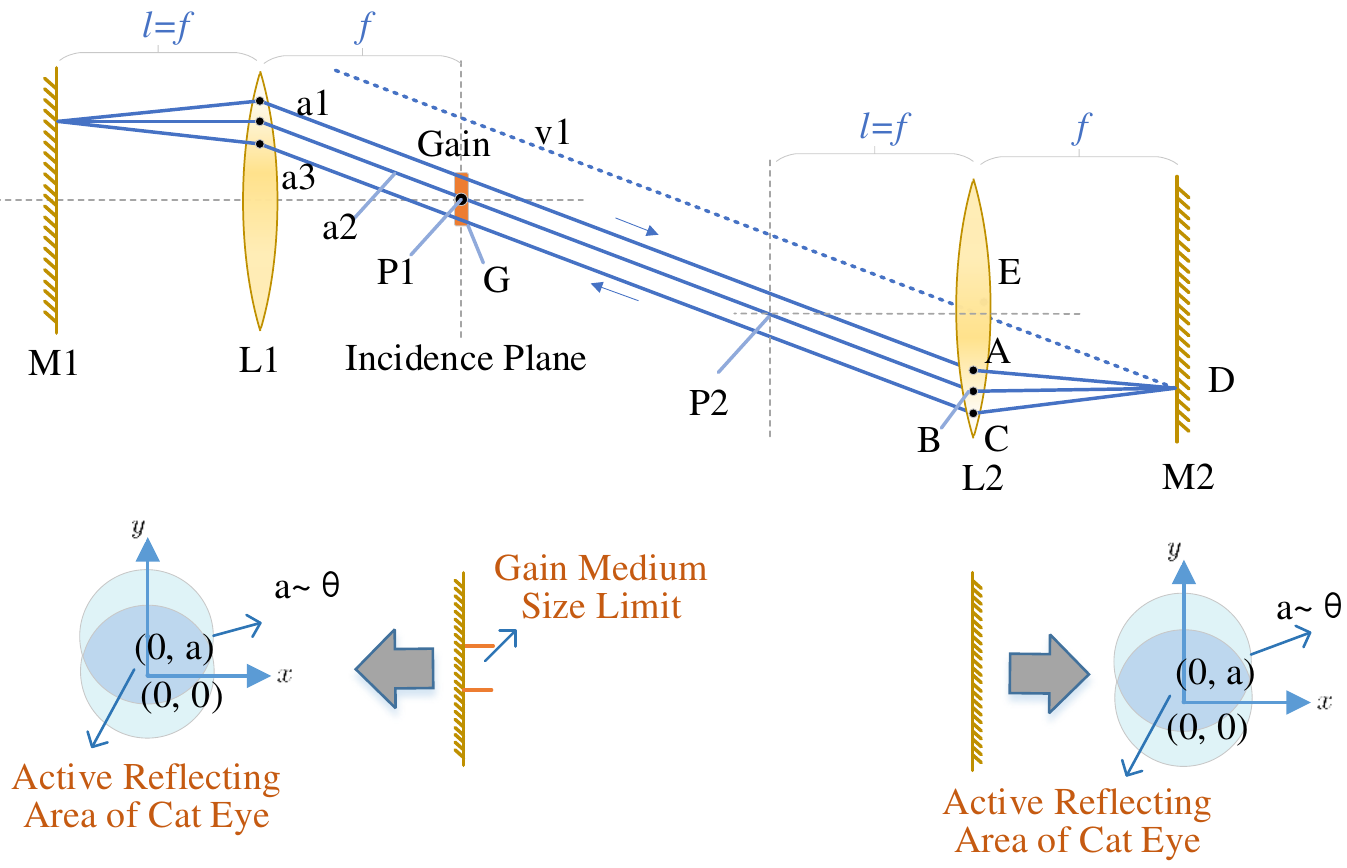}
    \caption{Equivalent resonator for mobile resonant cavity.}
    \label{f:Equivalency}
\end{figure}
To find out the energy distribution of the mobile transmission channel, we should at first obtain the output laser power $P_{\text{out}}$ in the receiver at any position within the system's FOV. As in Fig.~\ref{f:Equivalency}, P1P2 is the optic axis of the MRC. We place the diffraction limited gain medium at the focus of the left-hand cat's eye, to generate the resonant beam inside the MRC with the excitation of the pump power $P_{\text{in}}$. M2 is partially transmissive so that a portion of resonant beam will pass through M2 to form the laser beam with power of $P_{\text{out}}$. Following the laser principle, the relationship between $P_{\text{in}}$ and $P_{\text{out}}$ can be depicted as~\cite{hodgson2005laser}
\begin{equation}
\begin{aligned}
P_{\text{out}}=&A_{b} I_{S} \frac{(1-R)V_1 }{1-R V_1V_2+\sqrt{RV_1V_2 } \left[1 / (V_{S}V_1)-V_{S}\right]}\times\\
&\left[\frac{\eta_{\text{excit}}P_{\text{in}}}{A_gI_S}-\left| \mathrm{ln} \left(\sqrt{RV_{S}^2 V_1V_2} \right)\right|\right],
\label{e:outputpower2}
\end{aligned}
\end{equation}
where $A_g$, $I_s$, and $V_s$ are the cross sectional area, saturated light intensity, and loss factors of the gain medium, respectively; $\eta_{\text{excit}}$ is the excitation efficiency and $R$ is the reflectivity of M2; $A_b$ is the overlapping area of resonant beam and gain medium and $V_1,V_2$ (=1-loss) depict the loss factors during single intra-cavity transmission, respectively. To accurately obtain the $P_{\text{out}}$, we should calculate $V_{1,2}$ and $A_b$ exactly. At first, we employ the resonator mode analysis method to calculate the eigenmode of the MRC.

The active reflecting area of the cat's eye is affected by the incident angle $\theta$ of the beam. We adopt the indication function to depict
the reflective area by 1 and non-reflective area by 0. Then as in Fig.~\ref{f:Equivalency}, the active reflecting area of the left-hand and right-hand cat's eye can be depicted as
\begin{equation}
\begin{aligned}
T_1(x, y ; r, \theta)=\left\{\begin{array}{l}
1,  x^{2}+y^{2} \leq r^{2}, y \geq a / 2\\ \text { and } x^{2}+(y-a)^{2} \leq r^{2}, y \le a / 2 \\
0, \text { else }
\end{array}\right.
\end{aligned},
\end{equation}
where $a=2 f \tan \theta$, and $r$ is the radius of L1, L2, M1, and M2. Moreover, the diffraction limited gain medium limits the active reflecting area of the left-hand cat's eye, which can be depicted as
\begin{equation}
\begin{aligned}
T_2(x, y ; r_g, \theta)=\left\{\begin{array}{l}
1,  x^{2}+y^{2} \leq (r_g\cos{\theta})^{2}\\
0, \text { else }
\end{array}\right.
\end{aligned},
\end{equation}
where $r_g$ is the radius of the gain medium surface.

For the equivalent FP resonator in Fig.~\ref{f:Equivalency}, the active reflecting area of the right-hand mirror is $T_l = T_1\cdot T_2$, and $T_r = T_1$ for the right-hand mirror. Based on the above proof and inferences, we can calculate the modes of the FP resonator as of the MRC using diffraction theories.

\begin{figure}[t]
    \centering
    \includegraphics[width=3.5in]{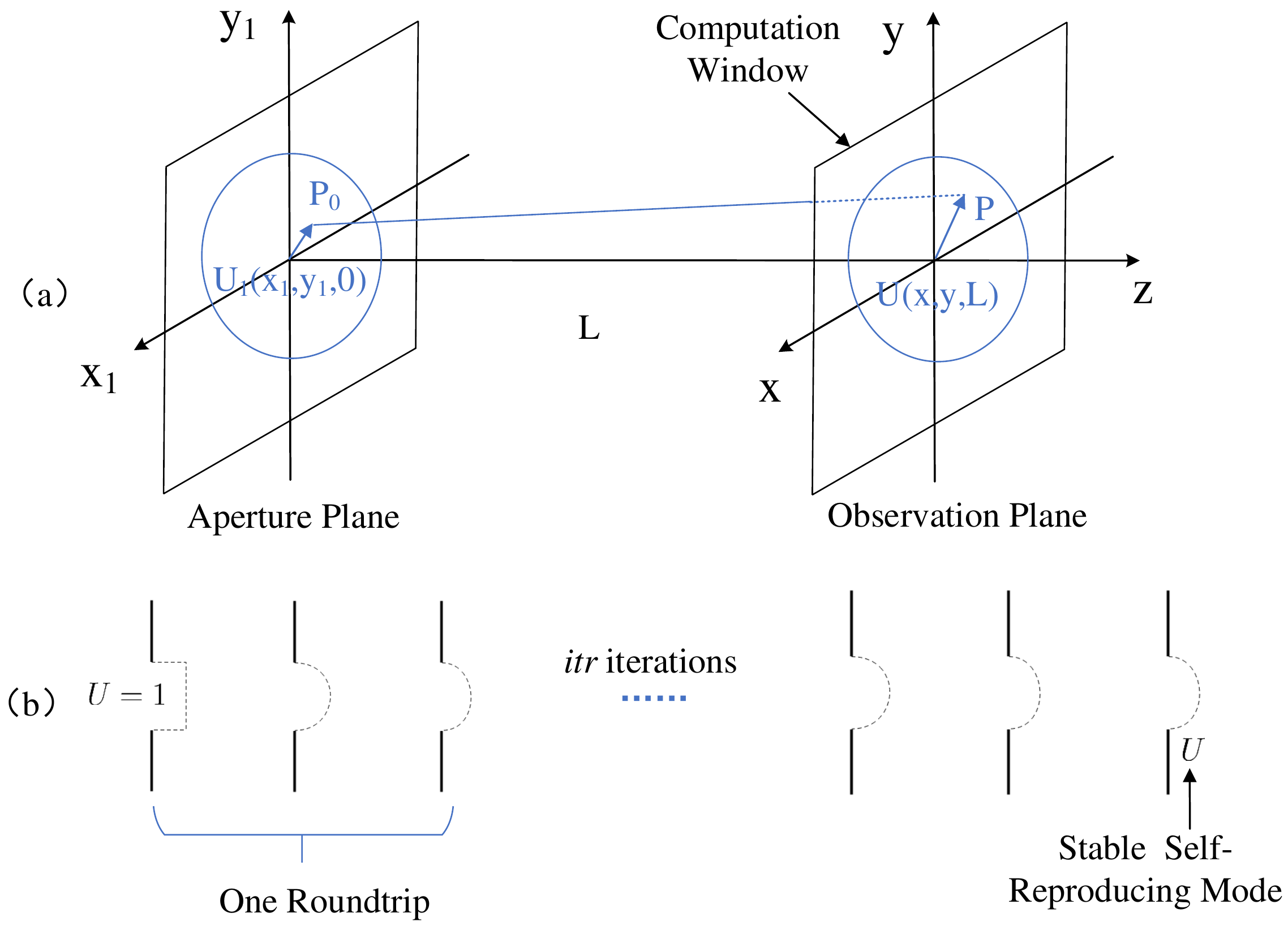}
    \caption{Illustration of diffraction theory and self-reproducing
mode calculation.}
    \label{f:diffraction}
\end{figure}
As in Fig.~\ref{f:diffraction} (a), if the light field distribution on one of the cavity mirror is known, one can obtain the amplitude and phase distribution of light field at any position in the cavity~\cite{Fresnel}. The Fresnel-Kirchhoff diffraction integral formula reads:
 \begin{equation}
 \label{e:FK}
 \begin{aligned}
U(x, y, L)=&\frac{\exp (j k L)}{j \lambda L} \iint_{T} U\left(x_{1}, y_{1}, 0\right)\\ &\exp \left\{j \frac{\mathrm{k}}{2 L}\left[\left(x-x_{1}\right)^{2}+\left(y-y_{1}\right)^{2}\right]\right\} \mathrm{d} x_{1} \mathrm{d} y_{1},
 \end{aligned}
\end{equation}
where $j:=\sqrt{-1}$, $L$ is the resonator length, $k=2\pi / \lambda$ is the wave number, and $\lambda$ is wavelength of resonant beam. Suppose $U(x_1, y_1, 0)$ represents the field on the left-hand mirror of the FP resonator, then $U(x, y, L)$ is the field on the other mirror and $T$ is the active reflecting area of the left-hand mirror.

For high-speed numerical calculation, Fast-Fourier-Transform (FFT) is adopted to find numerical solution of diffraction process~\cite{FFT}. \eqref{e:FK} can be rewritten as
the following convolution form:
\begin{equation}
U(x, y, L)= \iint_T U\left(x_{1}, y_{1}, 0\right) h\left(x-x_{1}, y-y_{1}\right) \mathrm{d} x_{1} \mathrm{d} y_{1} ,
\end{equation}
where $h\left(x-x_{1}, y-y_{1}\right) $ named as impulse response can be expressed as
\begin{equation}
h(x, y)=\frac{\exp (j k L)}{j \lambda z} \exp \left[\frac{j k}{2 L}\left(x^{2}+y^{2}\right)\right],
\end{equation}
then the formula can be rewritten as~\cite{FFT}
\begin{equation}
\mathscr{F}\{U(x,y,z)\}=\mathscr{F}\{U(x_1,y_1,0)\}\cdot \mathscr{F}\{h(x,y)\},
\end{equation}
where $\mathscr{F}$ denotes the Fourier transform. Moreover, the fields before and after passing through the cat's eye's front face $U^-(x,y)$ and $U(x,y)$ have the following relationship~\cite{shen2004mode}:
\begin{equation}
U(x,y)=T(x,y)\cdot U^-(x,y),
\label{e:Ut}
\end{equation}
where $T(x,y)$ represents the active reflecting area of the cat's eye. Thus, the self-consistent integral formula for one round-trip field transmission is
\begin{equation}
\begin{aligned}
U(x,y,L)=&\mathscr{F}^{-1}\{ \mathscr{F}\{\mathscr{F}^{-1}\{ \mathscr{F}\{U(x,y,L)\cdot T_{r}(x,y) \} \cdot \\
& \mathscr{F}\{h(x_1,y_1) \} \}\cdot T_{l}(x_1,y_1) \} \cdot \mathscr{F}\{h(x,y) \} \}\cdot \\
&T_{r}(x,y),
\end{aligned}
\label{e:selfconsisitent}
\end{equation}
where $\mathscr{F}^{-1}$ denotes the inverse Fourier transform process.

Based on Fox-Li method~\cite{FoxLi},
we are supposed to iterate \eqref{e:selfconsisitent} $t$ times to calculate the eigenmode of the resonator. As in Fig.~\ref{f:diffraction} (b), suppose the original light field distribution $U(x,y,L) = 1$, the light field on the output plane is gradually stable during iterations. Eventually, the forms of field distribution $U_t(x,y,L)$ after $t$ iterations and $U_{t-1}(x,y,L)$ after $t-1$ iterations are exactly identical, except for the constant factor difference on amplitude and phase. Then $U_t(x,y,L)$ can be regarded as the self-reproducing mode, or namely the eigenmode of the stable resonator.

\begin{figure}[t]
    \centering
        \subfigure[]{\includegraphics[width=3in]{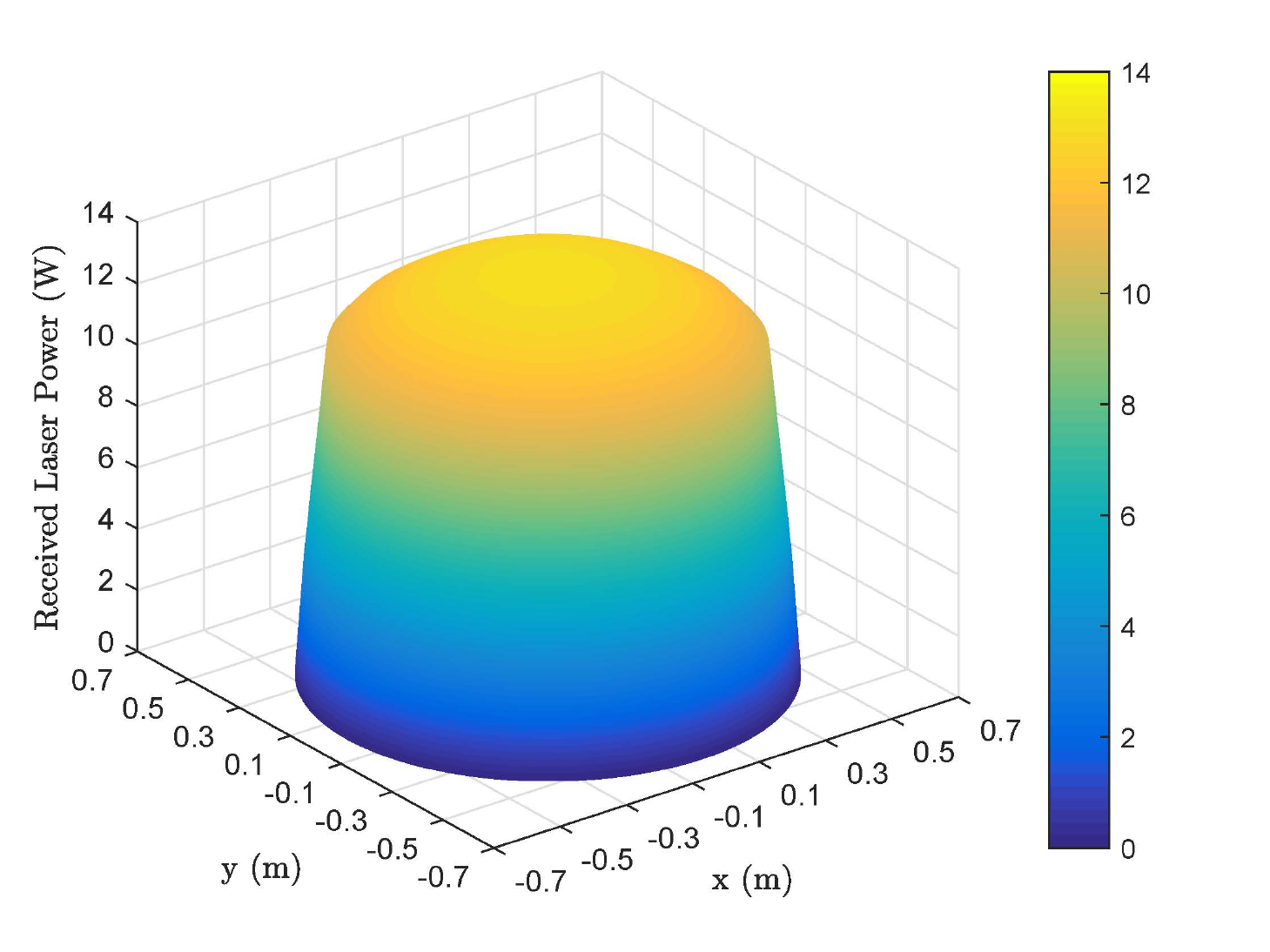}}\\ 
        \subfigure[]{\includegraphics[width=3in]{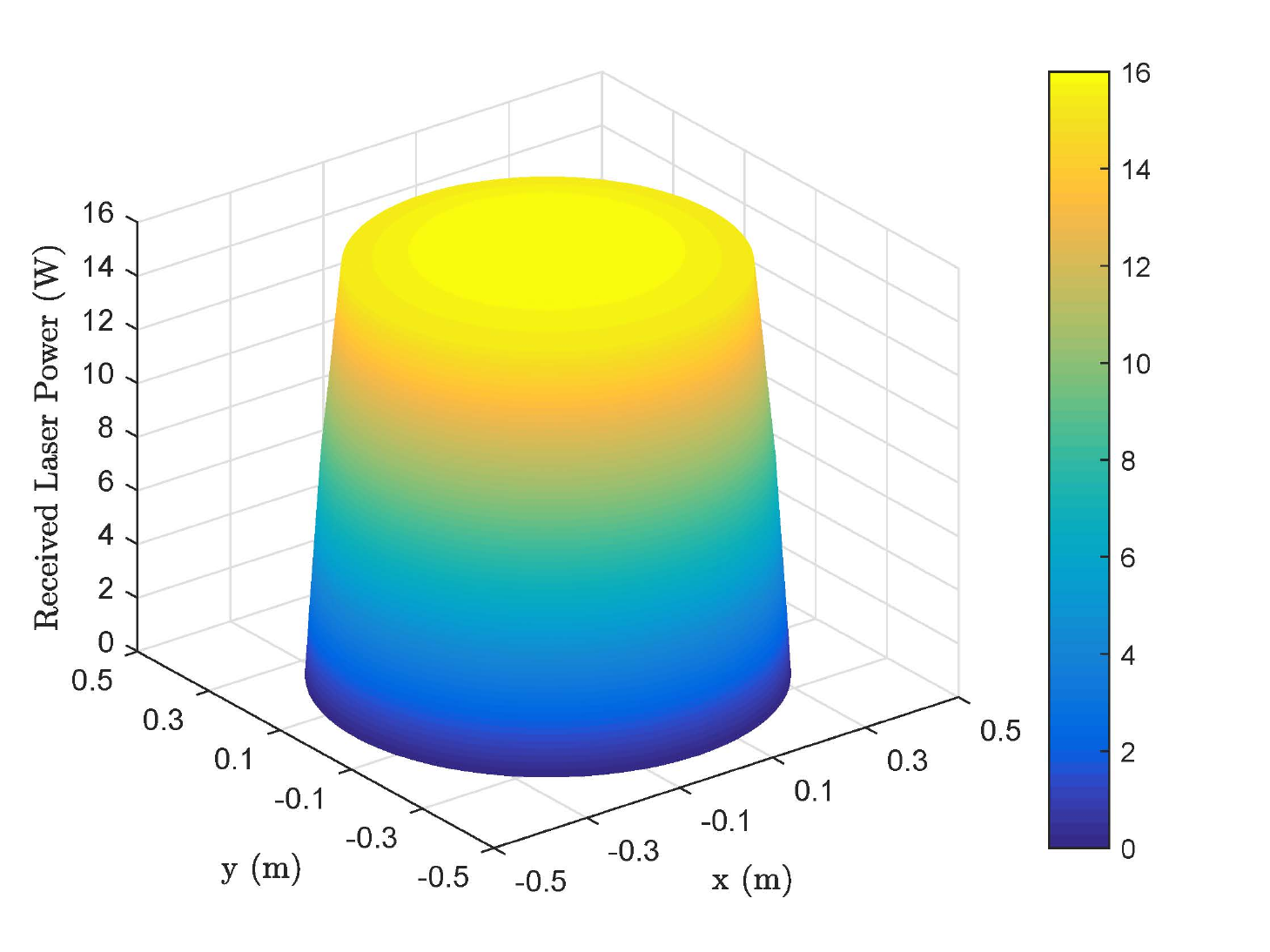}}
        \caption{Optical energy distribution (output laser power $P_{\text{out}}$ in the receiver) with the input power $P_{\text{in}}=150W$. The height of the transmitter is (a) 3 m; (b) 2 m. }
    \label{f:3dim}
\end{figure}

Then, we can find the beam spot size $A_s$ using the method for determining
Gaussian beam radius~\cite{Guassian} with $U_t(x,y,0)$, and the loss factor of the one-way transmission as
\begin{equation}
\label{e:lossfactor}
    V_1 = \frac{|U_t(x,y,L)|^2}{|U_{t}(x,y,0)|^2},\quad
    V_2 = \frac{|U_t(x,y,0)|^2}{|U_{t}(x,y,L)|^2}.
\end{equation}
Finally, we can obtain the receiver's output laser power $P_{\text{out}}$ with respect to the input power $P_{\text{in}}$, the distance $L$ between the two cat's eyes, and the translation angle $\theta$ of the remote cat's eye. Figure~\ref{f:3dim} depicts the power distribution on the receiving plane when the input power is $150$W and the receiving plane is $3$m and $2$m away from the transmitter.

\section{Energy and Data Flow in an exemplary RB-SLIPT system}
\label{sec:SLIPTFlow}
In this section, we present a practical RB-SLIPT system design embedding a PV panel for energy harvesting and an APD for data receiving respectively at the receiver. We will detail the energy and data transfer flow of RB-SLIPT from three aspects: signal generating and power input at the transmitter, path loss in the mobile transmission channel, and energy/data processing at the receiver.

\subsection{Signal Generating and Power Input at the Transmitter}
\subsubsection{Signal Generating}
We adopt the optical orthogonal frequency division multiplexing (OFDM) method for signal modulation~\cite{OFDM}. At the transmitter, let $s$ denote an input bit stream, the peak amplitude and variance of which can be depicted as $-A_{\text{c}} < i_{\text{sig}} < A_{\text{c}}$ and $\varepsilon$ respectively for $A_{\text{c}}>0$ and $\varepsilon > 0$. $s$ is at first mapped to the discrete modulation symbols, i.e., ${s_{1} ~ s_{2} ~ \cdots ~ s_{N / 2-1}}$, where $N$ is the number of subcarriers. Then, the Hermitian symmetry is imposed on the data vector as $\mathbf{S} = \left[
0 ~ s_{1} ~ s_{2} ~ \ldots ~ s_{N / 2-1} ~ 0 ~ s_{N / 2-1}^{*} ~ \cdots ~ s_{2}^{*}
~ s_{1}^{*}\right]$ so that the output of inverse discrete
Fourier transform (IDFT) block can be real~\cite{gamma}.
As illustrated in Fig.~\ref{f:overview}, we denote $i_{\text{sig}}(t)$ as a source signal from the electrical modulator; then $i_{\text{sig}}(t)$ after the IDFT operation can be presented as
\begin{equation}
i_{\text{sig}}(t)=\sum_{k=0}^{N-1} \underbrace{\frac{1}{\sqrt{N}} S(k) e^{j \frac{2 \pi k}{N} t}}_{i_{\text{sig}, k}(t)}, t=0,1, \ldots, N-1,
\end{equation}
where $i_{\text{sig}, k}(t)$ represents the signal on $k_{th}$ subcarrier, and $S(k)$ is the corresponding element of $\mathbf{S}$.

Then, a DC component from the pump source acts as a DC offset to the source signal. Thus, the generated signal at the transmitter can be given as
\begin{equation}
I_{\text{in}}(t) = I_{\text{D}} + i_{\text{sig}}(t),
\label{e:Iin}
\end{equation}
where $I_{\text{D}}$ is the DC offset and $i_{\text{sig}}$ is an AC component which carries the data that needs to be sent out.


\subsubsection{Power Input}
For a thin-disk solid laser, an LD is adopted to pump the thin-disk gain medium. The electric current $I_{\text{in}}$ at first drives the LD to generate the pump laser with power $P_{\text {pump}}$ with the relationship as
\begin{equation}
\label{e:pump}
\begin{aligned}
P_{\text {pump }}=\frac{h c}{q \lambda_{\text{e}}} \eta_{\nu}\eta_{e}\left[I_{\text {in }}-I_{\text {th }}\right]
\end{aligned}
\end{equation}
where $h=6.62607015\times 10^{-34}\text{J}\cdot\text{s}$, $c=3\times 10^8 \text{m/s}$, $q=1.6 \times 10^{-19}\text{C}$, $\lambda_{\text{e}} = 808\text{nm}$ are the Planck’s constant, the speed of light, the electron charge, and the emission light wavelength, respectively; $\eta_{\nu}$ and $\eta_{e}$ are carrier injection efficiency and photon extraction efficiency; and $I_{\text{th}}$ is a temperature-dependent constant threshold current. $I_{\text{D}}$ contributes to the input current exceeding the LD's threshold current $I_{\text{th}}$.

Then, the pump power is absorbed by the gain medium and converted to the power stored in the upper laser level inside the gain medium $P_{\text {avail}}$ with efficiency $\eta_a$ as
\begin{equation}
\label{e:available}
\begin{aligned}
P_{\text {avail}}&=\eta_a P_{\text {pump}}\\
&=\eta_{\text {excit}} P_{\text {in}}
\end{aligned},
\end{equation}
where $\eta_{\text {excit}} = \eta_{\text{a}}\eta_{\text{P}}$ represents the excitation efficiency. $P_{\text {in}}$ denotes the input power to the laser diode, and $\eta_{\text{P}}$ is defined as the pump efficiency. Thus, $P_{\text {avail}}$ is available for generating the resonant beam, which can be regarded as the power of the transmitted signal.

\subsection{Path Loss in the Mobile Transmission Channel}
For traditional SLIPT systems, the path loss of transmission channel is caused by the beam divergency and misalignment of the transceiver. For RB-SLIPT system, the resonant beam power which carries information and energy will increase as passing through the gain medium, and decrease as experiencing the diffraction losses, losses inside the medium, and output coupling. Due to the characteristics of long-range intra-cavity transfer and self-alignment, the intra-cavity diffraction loss cannot be neglected in the mobile transmission channel. As in Sec III, we can obtain the diffraction loss of one-way transmission inside the MRC as \eqref{e:lossfactor}. Thus, the diffraction loss factor $\eta_{\text{diff}}$ can be depicted as
\begin{equation}
\label{e:diff}
\eta_{\text{diff}} = \sqrt{V_1V_2}.
\end{equation}

In the RB-SLIPT system, the diffraction loss is 	dominant to the path loss during long-range transmission. From Sec III, the diffraction loss in the mobile transmission channel relies on the size of system components, and the distance and deflection angle between the transceiver.

\subsection{Energy/Data Processing at the Receiver}
Similar to the traditional laser system, the output laser power $P_{\text{out}}$ will be extracted from $P_{\text{avail}}$ after experiencing the path loss during the channel transmission, internal loss of the gain medium, and the loss due to the output coupling, where the extraction efficiency can be depicted as~\cite{hodgson2005laser}
\begin{equation}
\eta_{\text{extr}}= \frac{\eta_{\text{b}}(1-R)V_1 }{1-\eta_{\text{diff}}^2R +\eta_{\text{diff}}\sqrt{R} \left[1 / (V_{S}V_1)-V_{S}\right]},
\label{e:extraction}
\end{equation}
with $\eta_{\text{b}} = A_{\text{b}}/A_{\text{g}}$ denoting the overlap efficiency of the gain medium.

Then, the output laser power can be presented from \eqref{e:outputpower2}, \eqref{e:available}, \eqref{e:diff}, and \eqref{e:extraction} as
\begin{equation}
P_{\text{out}}=\eta_{\text{extr}}(\eta_{\text{excit}}P_{\text{in}}-c_1),
\label{e:simplyoutput}
\end{equation}
where $c_1 = A_gI_s\left| \mathrm{ln} \left(\sqrt{RV_{S}^2 V_1V_2} \right)\right|$ represents the threshold of laser output in the RB-SLIPT system. The parameter $\eta_{\text{extr}}$ contains the impacts of various losses in the system, and $\eta_{\text{excit}}$ depicts the stimulation process of the thin-disk gain medium. Then, given a split ratio $0\le \mu\le 1$, $\mu P_{\text{out}}$ is received by a PV panel for energy harvesting and $(1-\mu)P_{\text{out}}$ is collected by an APD for information decoding.

\subsubsection{Energy Harvesting}
\begin{figure}[t]
    \centering
    \includegraphics[width=2.5in]{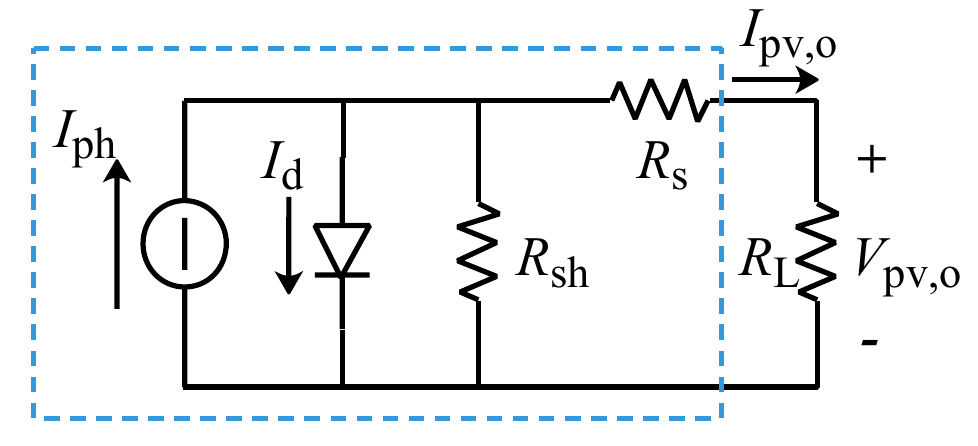}
    \caption{DC Equivalent Circuit of PV for Energy Harvesting}
    \label{f:DCPV}
\end{figure}

Figure~\ref{f:DCPV} depicts the equivalent circuit of PV for energy harvesting. PV under illumination can be regarded as a constant current source in parallel with diode. $I_{\text{ph}}$ represents the photo-generated current, and $I_{\text{d}}$ denotes the current that used to counteract the junction current of p-n junction. The additional resistance due to PV's material properties is represented by series resistance $R_{\text{s}}$ and
the edge leakage is modeled by a parallel shunt resistance $R_{\text{sh}}$. Finally, $R_{\text{L}}$ represents the load, i.e., battery to be charged, with the PV's output current $I_{\text{pv,o}}$ and output voltage $V_{\text{pv,o}}$.

Then according to Kirchhoff’s law, the current-voltage (I-V) characteristics of a PV panel at a maximum power point can be described as
\begin{equation}
\label{e:pv1}
I_{\mathrm{pv}, \mathrm{o}}=I_{\mathrm{ph}}-I_{\mathrm{d}}-\frac{V_{\mathrm{d}}}{R_{\mathrm{sh}}},
\end{equation}
where $V_{\text{d}}$ is the voltage across the diode follows
\begin{equation}
\label{e:pv2}
V_{\mathrm{d}}=V_{\mathrm{pv}, \mathrm{o}}+I_{\mathrm{pv}, \mathrm{o}} R_{\mathrm{s}}.
\end{equation}
Moreover, the current $I_{\text{d}}$ through the diode is
\begin{equation}
\label{e:pv3}
I_{\mathrm{d}}=I_{0}\left(\mathrm{e}^{c_{\text{1}}{V_{\mathrm{d}}}}-1\right),
\end{equation}
where $I_{\text{0}}$ is the reverse saturation current, $c_{\text{1}} = 1/ {n_{\mathrm{s}} n V_{T}}$ is the PV panel factor, with $n_{\text{s}}$ the number of cells connected inside a PV panel, $n$ the diode ideality factor, and $V_{T}=(k T)/{q}$ is the thermal voltage of the diode, with $k$ the Boltzmann's constant and $T$ the temperature in Kelvin.

Moreover, $I_{\text{ph}}$ depends on the light power received by the PV panel $P_{\text{pv,i}}$. Assume the output laser power for energy harvesting is 100\% harvested by the PV, i.e., $P_{\text{pv,i}} = \mu P_{\text{out}}$, the relationship between photocurrent $I_{\text{ph}}$ and PV's input laser power $P_{\text{pv,i}}$ can be depicted as
\begin{equation}
\label{e:pv3}
\begin{aligned}
I_{\text{ph}} &= \rho_1 P_{\text{pv,i}}\\
&=\mu\rho_1\eta_{\text{excit}}(\eta_{\text{extr}}P_{\text{in}}-c_1),
\end{aligned}
\end{equation}
where $\rho_1$ is the conversion responsivity factor which depicts the optical-to-electrical conversion efficiency and can be measured in A/W. From \eqref{e:pv1}, \eqref{e:pv2},
and \eqref{e:pv3}, we can obtain the output power $P_{\text{pv,o}}$ of the PV panel available for charging the battery as
\begin{equation}
P_{\text{pv,o}} = I_{\text{pv,o}}V_{\text{pv,o}}.
\end{equation}

\subsubsection{Data Receiving}
A portion of laser beam is received by the APD to be converted to the output signal current as
\begin{equation}
P_{\text{pd}} = (1-\mu)P_{\text{out}},
\label{e:APD}
\end{equation}
At the receiver part, there are three following aspects need to be detailed for the communication performance analysis.
\begin{itemize}
\item Received Signal: To present the signal gain in the mobile transmission channel of the RB-SLIPT, we can rewrite \eqref{e:APD} as follows according to \eqref{e:pump} and \eqref{e:simplyoutput}:
\begin{equation}
i_{\text{pd}} = \sqrt{1-\mu}\left(\gamma \left[I_{\text {in }}-I_{\text {th }}\right] + c_2\right),
\label{e:ipdo}
\end{equation}
where $\rho_2$ is a constant representing the optical-to-electrical conversion responsivity of APD, and $\gamma$ and $c_2$ can be presented as follows:
\begin{equation}
\gamma=\rho_2\eta_{\text{extr}}\eta_{\text{a}} \eta_{\nu}\eta_{e}\frac{h c}{q \lambda_{\text{e}}}.
\label{e:gamma}
\end{equation}
$\gamma$ and $c_2$ are constants at given transmission distance and deflection angle between the transceiver, which illustrates the linear modulation ability of mobile transmission channel for communication. Then, $\gamma$ is modelled as the channel gain for signal transfer in RB-SLIPT, which reflects the channel condition depending on the transmission distances and deflection angles. Thus, the time domain signal on $k_{th}$ subcarrier received by the user can be described as~\cite{OFDMsplit}
\begin{equation}
i_{\text{pd},k}(t) = \sqrt{1-\mu}\gamma i_{\text{sig},k}(t) + n_k(t),
\end{equation}
where $n_k(t)$ is the noise signal on the $k_{th}$ subcarrier.
\item Noise Analysis: We analyze the shot noise generated through the optical-to-electrical conversion in the APD and the thermal noise caused by the resistors in the APD system. The variance of the noise signal with zero mean on $k_{th}$ subcarrier can be expressed as~\cite{gamma}
    \begin{equation}
    \sigma_k^2 = N_{\text{total}}W/N,
    \end{equation}
    where $W$ is the modulation bandwidth. The noise signal can be modelled as additive white Gaussian noise with power spectral density (PSD) profile denoted as $N_{\text{total}}$, which is the sum of shot noise and thermal noise:
    \begin{equation}
    N_{\text{total}} = N_{\text{sh}} + N_{\text{th}}.
    \end{equation}

    The one-side PSD of the shot noise in $\text{A}^{\text{2}}/\text{Hz}$ can be depicted as~\cite{Xiong2}
    \begin{equation}
    \begin{aligned}
    N_{\text{sh}} = 2q\rho_2 (P_{\text{pd}} + P_{\text{b}}),
    \end{aligned}
    \end{equation}
    where $P_{\text{b}}$ is the background radiance caused by the ambient light (sunlight or LED light), which can be depicted as~\cite{Pbck}
    \begin{equation}
    P_{\mathrm{b}}=\eta_{\mathrm{Rx}} H_{\mathrm{b}} B_{\mathrm{IF}} A_{\mathrm{Rx}} \Phi_{\mathrm{Rx}} \Gamma,
    \end{equation}
    where $\eta_{\mathrm{Rx}}$ is the optical efficiency of the signal receiving unit, $H_{\mathrm{b}}$ represents the background irradiance, $B_{\mathrm{IF}}$ is the optical bandwidth of the filter installed behind the output coupler retro-reflector2; $A_{\mathrm{Rx}}$, $\Phi_{\mathrm{Rx}}$, and $\Gamma$ are the receiving area, solid angle, and transmittance of retro-reflector2, respectively. In our settings, the background irradiance is estimated as $9.56\times 10^{-6}$W.

    The thermal noise is generated by the load resistor $R_{\text{L,pd}}$, of which the one-side PSD can be described as~\cite{totalNoise}
    \begin{equation}
        N_{\text{th}}=4 k T / R_{\mathrm{L,pd}}.
    \end{equation}

\item Signal-to-Noise Ratio (SNR): To evaluate the performance of the communication in the RB-SLIPT system, we analyze the SNR of the channel. Due to the OFDM method, we can choose appropriate modulation scheme for each sub-channel and design the optimized bit distribution algorithm, so that the utilization of the channel capacity can be maximized. For simplicity, we depict the channel performance from the SNR of each sub-channel, and the SNR of the $k_{th}$ sub-channel can be~\cite{gamma}
    \begin{equation}
    \mathrm{SNR}_{k}=\frac{(1-\mu)\gamma^{2}E_k}{\sigma_k^2},
    \end{equation}
    where $E_k$ represents the power of transmitted signal $i_{\text{sig},k}$ on $k_{th}$ subcarrier, which can be estimated as~\cite{gamma}
    \begin{equation}
    E_{ k}=\frac{\left(E\left[i_{\text{sig},k}(t)\right]\right)^{2}}{\eta^{2}(N-2)} = \frac{2\varepsilon}{\eta^{2}(N-2)}.
    \end{equation}
    It should be noted that only $N-2$ subcarriers carry data and $\eta$ denotes the DC bias factor.

    Then, according to the Shannon channel capacity, the achievable data rate of the RB-SLIPT system for communication can be depicted as~\cite{gamma}
    \begin{equation}
    R_{\text{a}}=\sum_{k=1}^{N / 2-1}(W / N) \log _{2}\left(1+\mathrm{SNR}_{k}\right).
    \end{equation}
\end{itemize}

\section{Numerical Analysis}
In this section, we numerically analyzed the channel factor and the energy/data transfer capability of the RB-SLIPT system. As for the mobility-enhanced system, we will specifically analyze the impacts of the moving range ( i.e., transmission distance and deflection angle), and analyze the impacts of the DC offset power and the split ratio on the performance of the proposed system.

\subsection{Parameters}

\begin{table}[!htbp]
\newcommand{\tabincell}[2]{\begin{tabular}{@{}#1@{}}#2\end{tabular}}
\centering
\caption{~Parameter of Resonant Beam System}
\vspace{.7em}
\begin{tabular}{ccc}
\hline
\textbf{Parameter}&\textbf{Symbol}&\textbf{Value}\\
\hline
\text{Cat's radius}&$r$&$12$mm\\
\text{Gain medium radius}&$r_g$&$3$mm\\
\text{Resonant beam wavelength}&$\lambda$ & $1.064\times10^{-6}$m\\
\text{Output coupler reflectivity}&$R$&$0.7$\\
\text{Medium saturated intensity}&$I_s$&$1.26\times10^{7}$W/m$^2$\\
\text{Loss factor in medium}&$V_s$&$0.99$\\
\text{Excitation efficiency}&$\eta_{\text{excit}}$&$0.5148$\\
\text{Gain stored efficiency}&$\eta_{\text{a}}$&$0.72$\\
\tabincell{c}{\text{Carrier injection and}\\ \text{photon extraction efficiency}}&$\eta_{\nu}\eta_{e}$&$0.715$\\
\hline
\label{t:systemPara}
\end{tabular}
\end{table}

\begin{table}[!htbp]
\centering
\caption{~Parameter of Energy/Data Transfer}
\vspace{.7em}
\begin{tabular}{ccc}
\hline
\textbf{Parameter}&\textbf{Symbol}&\textbf{Value}\\
\hline
\text{PV conversion responsivity}&$\rho_1$&$0.746$A/W\\
\text{Reverse saturation current}&$I_\text{0}$&$9.381\times 10^{-9}$A\\
\text{Diode ideality factor}&$n$ & $1.318$\\
\text{Number of PV cell}&$n_{\text{s}}$ & $1$\\
\text{Series resistance}&$R_{\text{s}}$ & $25$m$\Omega$ \\
\text{Shunt resistance}&$R_{\text{sh}}$ & $5000\Omega$\\
\text{PD load resistor}&$R_{\text{L}}$ & $25$m$\Omega$ \\
\text{Temperature in Kelvin}&$T$ & $298.15$K\\
\text{APD conversion responsivity}&$\rho_2$&$0.6$A/W\\
\text{Modulation bandwidth}&$W$&$200$M/Hz\\
\text{PD load resistor}&$R_{\text{L,pd}}$&$10$K$\Omega$\\
\text{DC bias factor}&$\eta$&$3$\\
\hline
\label{t:SWIPTPara}
\end{tabular}
\end{table}

\begin{table}[!htbp]
\centering
\caption{~Parameter of FoxLi and FFT algorithm}
\vspace{.7em}
\begin{tabular}{ccc}
\hline
\textbf{Parameter}&\textbf{Symbol}&\textbf{Value}\\
\hline
\text{FoxLi iteration number}&$t$&$300$\\
\text{Sampling number}&$S_N$&$2048$\\
\text{Computation window expand factor}&$G$ & $2$\\
\hline
\label{t:FFTPara}
\end{tabular}
\end{table}

The parameters of the gain medium are from an example of a diode-end-pumped Nd:YVO4 laser  in~\cite{hodgson2005laser,Excitation}. We choose the infrared beam with the wavelength of $1064$nm. Then, we adopt a PV cell with $0.746 $A/W and an APD with $0.6 $A/W conversion responsivity~\cite{gamma}.
We specify the RBS parameters, energy and data transfer parameters, and FFT and FoxLi algorithm parameters in Table~\ref{t:systemPara}, ~\ref{t:SWIPTPara}, and  ~\ref{t:FFTPara}, respectively.

For the FFT numerical calculation, the calculation plane and the impulse response function need to be sampled first, and the sampling number $S_N$ is generally a power of $2$. Moreover, zero-padding is needed to avoid aliasing effects as in Fig.~\ref{f:diffraction}. The computation window length $2Gr$ is defined as the length of zero-padded aperture, where $G$ is the computation window expand factor, and $r$ is the front face radius of cat's eye.

\subsection{Channel Factor of the Mobile Transmission Channel}

\begin{figure}[t]
    \centering
    \includegraphics[width=3.5in]{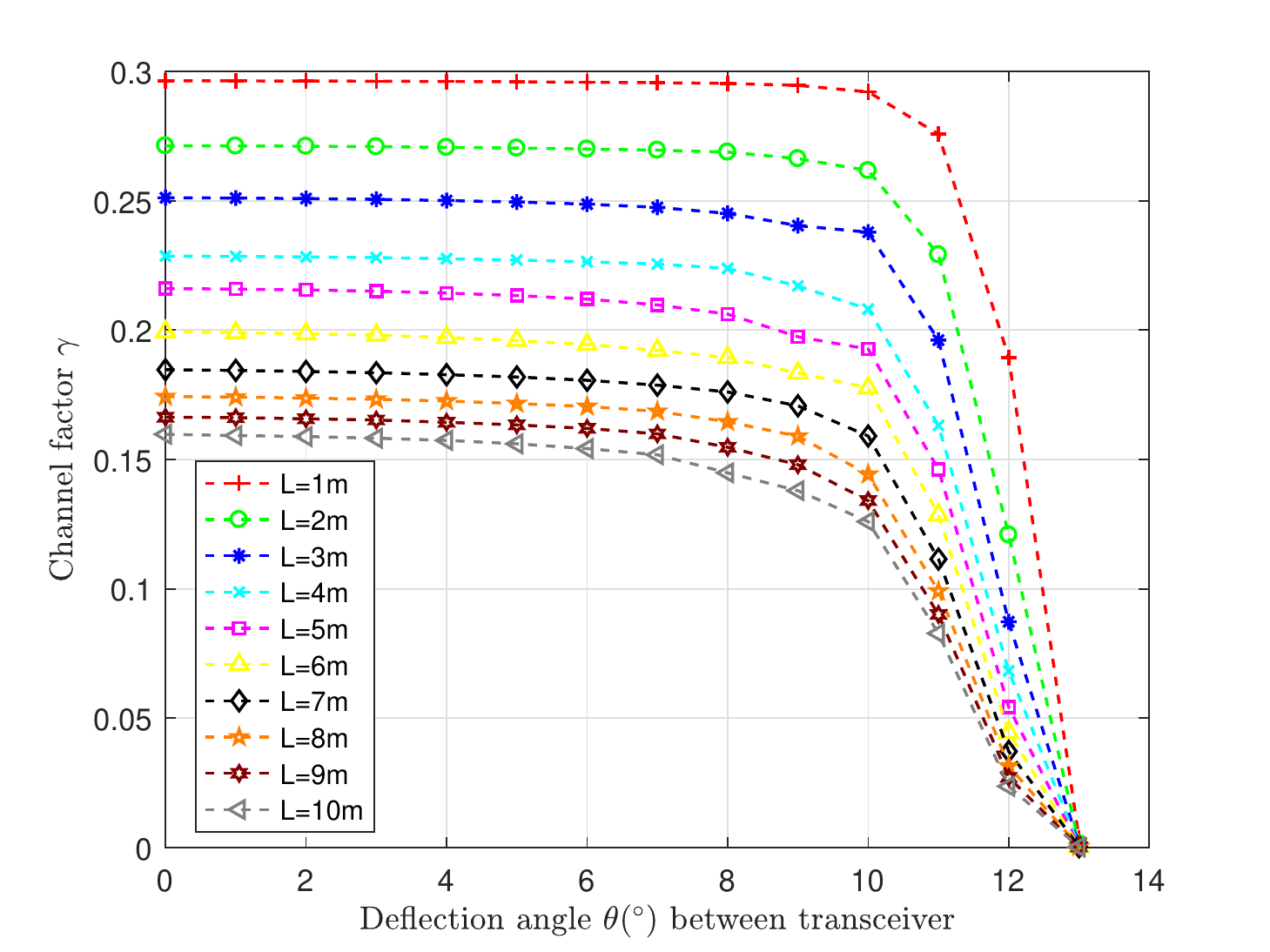}
    \caption{Channel factors $\gamma$ of the RB-SLIPT with different transmission distances $L$ and deflection angles $\theta$.}
    \label{f:gamma}
\end{figure}
In this paper, we have proposed an analytical model for the energy distribution analysis of the mobile transmission channel, so that we are capable of modelling the channel accurately as a function of mobility factors such as transmission distance, i.e., cavity length $L$ and the deflection angle between transceiver $\theta$. Figure~\ref{f:gamma} depicts the channel factors $\gamma$ of the proposed system with varying $L$ and $\theta$. $\gamma$ doesn't change obviously as the deflection angle between the transceiver is below $10^{\circ}$. However, as the angle is larger than $10^{\circ}$, $\gamma$ declines sharply and turns to $0$ with $\theta = 13^{\circ}$. The maximum allowable angle depends on the parameters (i.e., focus length and front face radius) of the cat's eye. Moreover, as the transmission distance $L$ changes from $1$m to $10$m, $\gamma$ decreases correspondingly.

\subsection{Energy-Data Transfer Performance of RB-SLIPT}

\begin{figure}[t]
    \centering
    \includegraphics[width=3.5in]{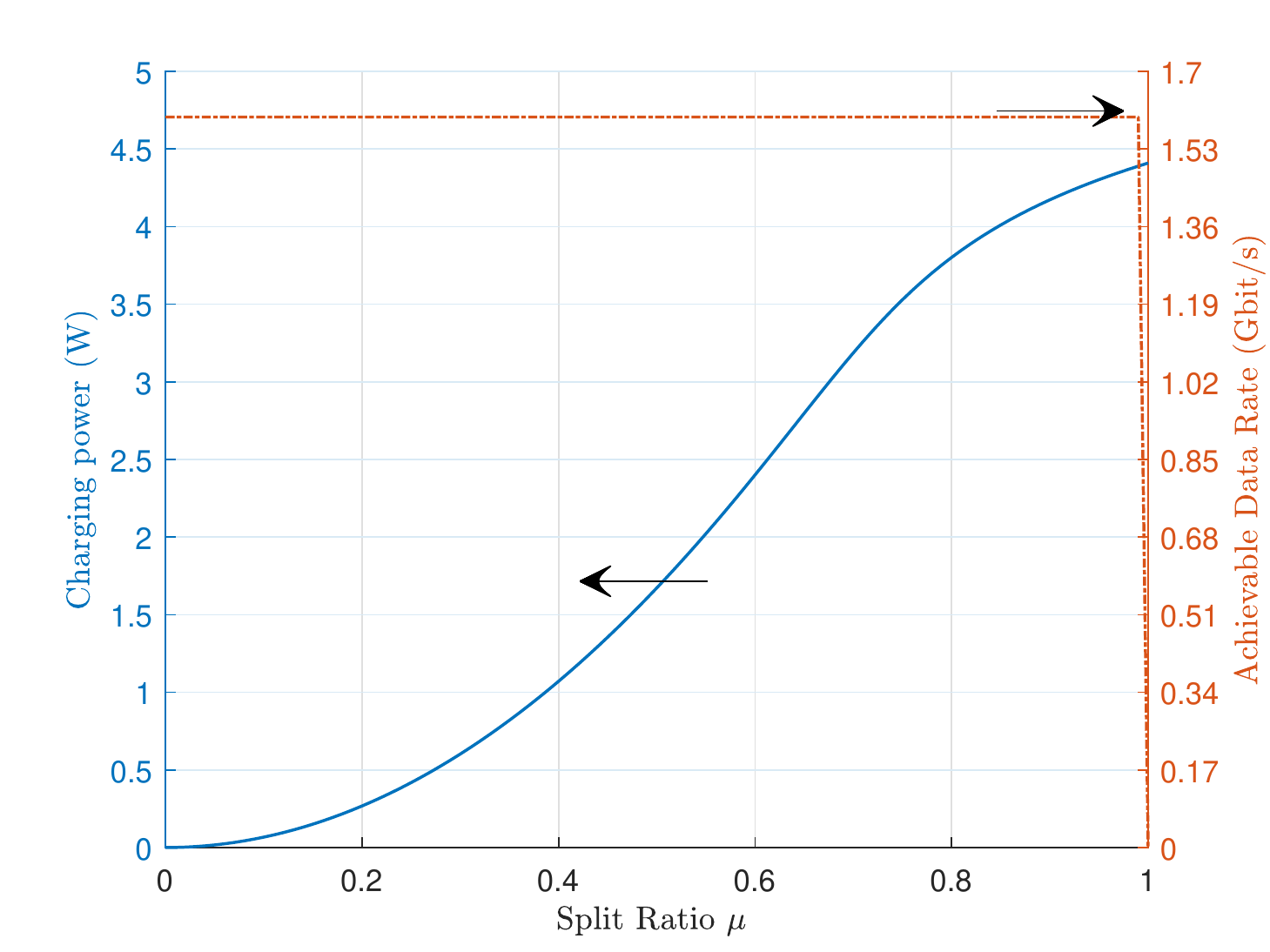}
    \caption{Charging power and achievable data rate as a function of deflection angle $\theta$ with different split ratio $\mu$ under transmission distance $L=3$m.}
    \label{f:responses}
\end{figure}
At the receiver, the output laser beam at first enters into a power splitter and is split into two streams with ratio $\mu$ for energy and data harvesting, respectively. Thus, as in Fig.~\ref{f:responses}, we analyze how the split ratio impacts the energy/data transfer performance. We can find as $\mu$ grows, the charging power improves significantly, while there is no obvious deterioration in data transfer performance until $\mu = 1$. Thus, in the following analysis, we choose the split ratio $\mu = 0.99$ for maximizing the energy/data transfer capability.

\begin{figure}[t]
    \centering
    \includegraphics[width=3.5in]{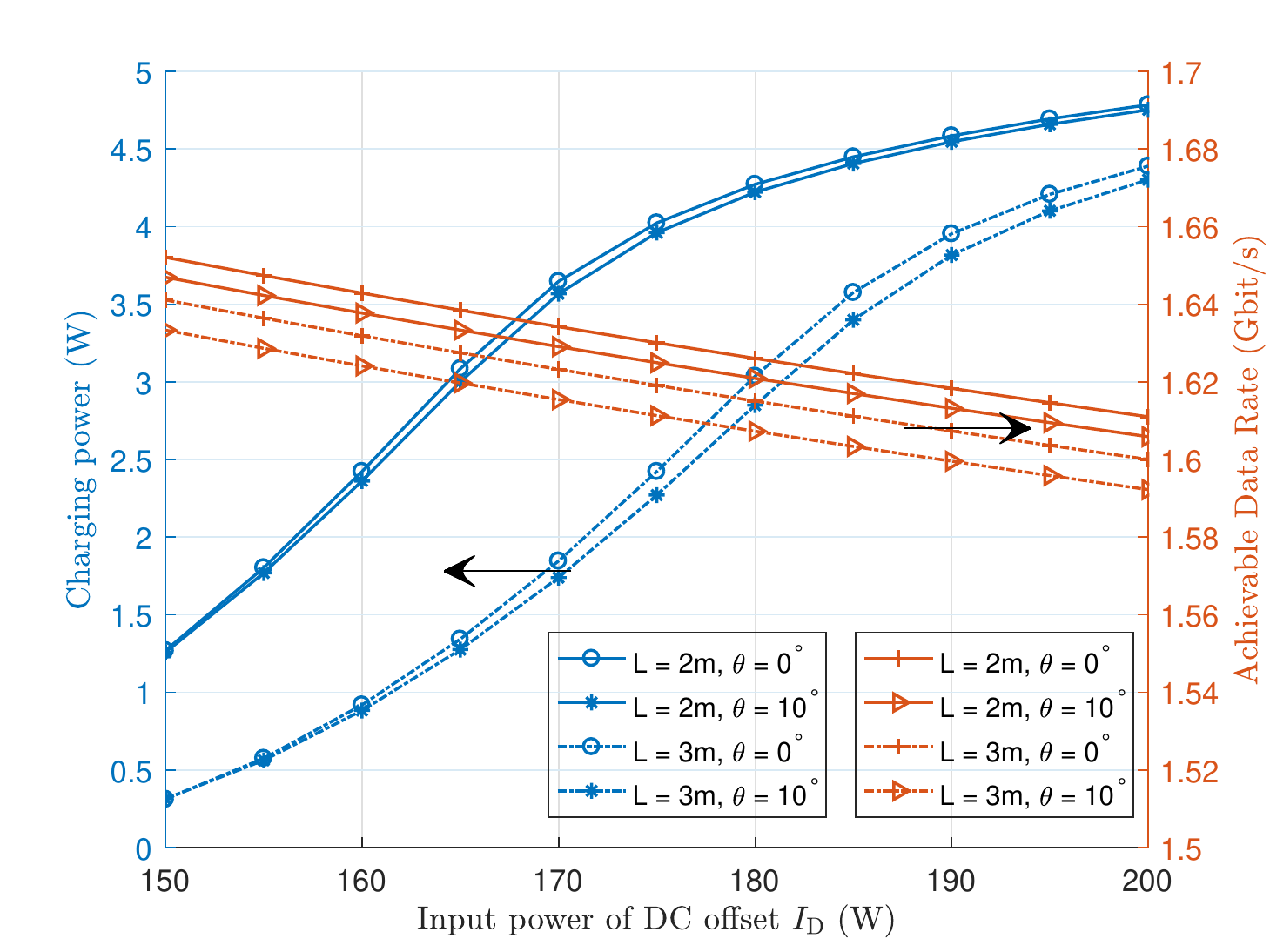}
    \caption{Charging power and achievable data rate as a function of input power of the DC offset $I_{\text{D}}$ with different transmission distances and deflection angles.}
    \label{f:Pin}
\end{figure}
Figure~\ref{f:Pin} shows the charging power and achievable data rate as a function of input power of the DC offset $I_{\text{D}}$, with transmission distance $L=2$m, $3$m and deflection angle $\theta = 0^{\circ}, 10^{\circ}$. As the input power increases, the charging power from PV grows, while the communication performance gets poorer due to the increase of noises.

\begin{figure}[t]
    \centering
    \includegraphics[width=3.5in]{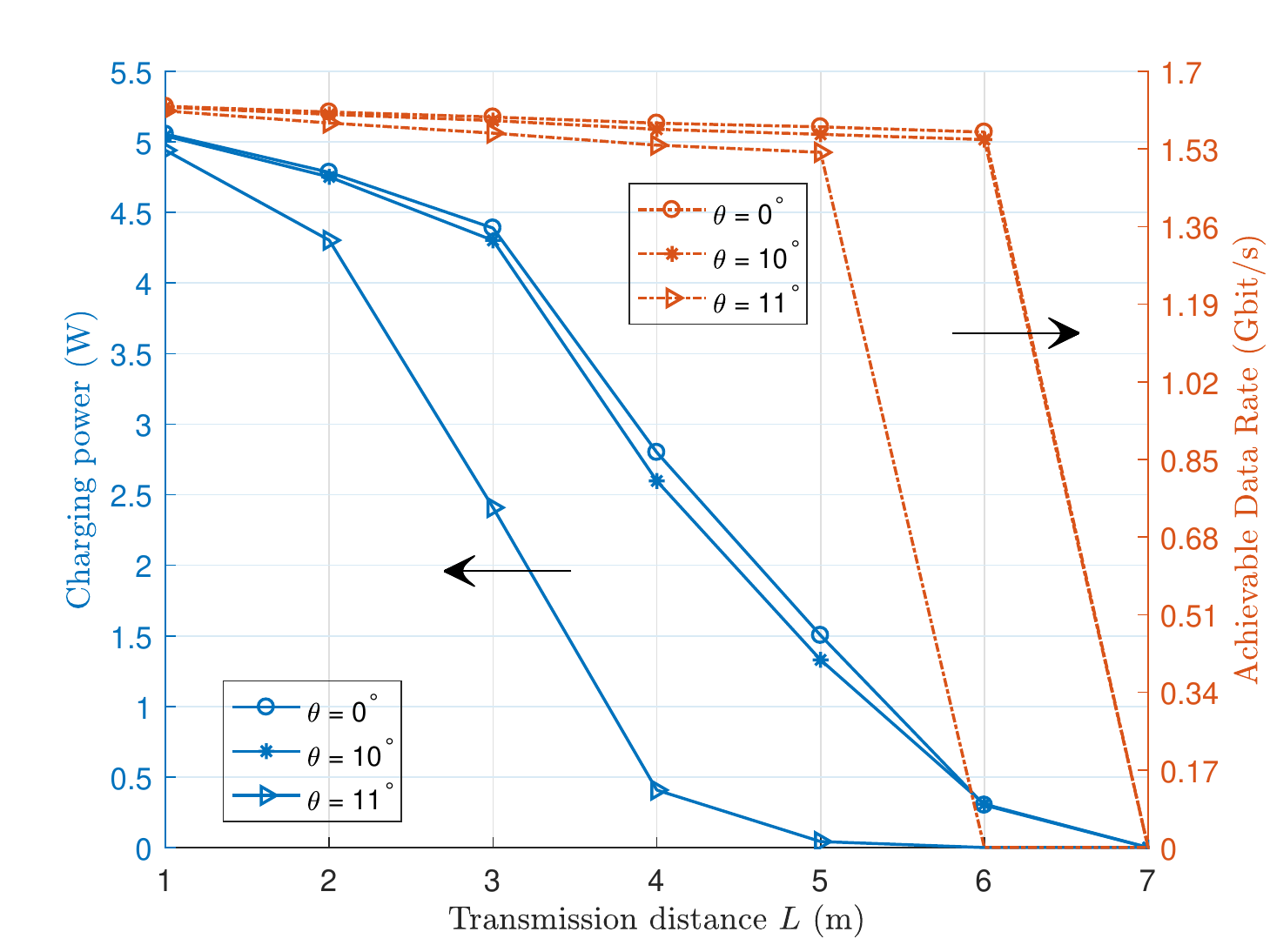}
    \caption{Charging power and achievable data rate as a function of transmission distance $L$ with different deflection angles under input power $P_{\text{in}} = 200$W.}
    \label{f:L}
\end{figure}
Figure~\ref{f:L} depicts the charging power and achievable data rate as a function of transmission distance $L$ with the deflection angle $\theta = 0^{\circ}, 10^{\circ}, 15^{\circ}$, and the input power to the system $P_{\text{in}} = 200$W. Evidently, both the energy and data transfer performances reduce with the increase of the transmission distance $L$. Moreover, $L$ has more impact on the energy transfer than on the data transfer, and as the deflection angle grows, the acceptable transmission distance decreases.

\begin{figure}[t]
    \centering
    \includegraphics[width=3.5in]{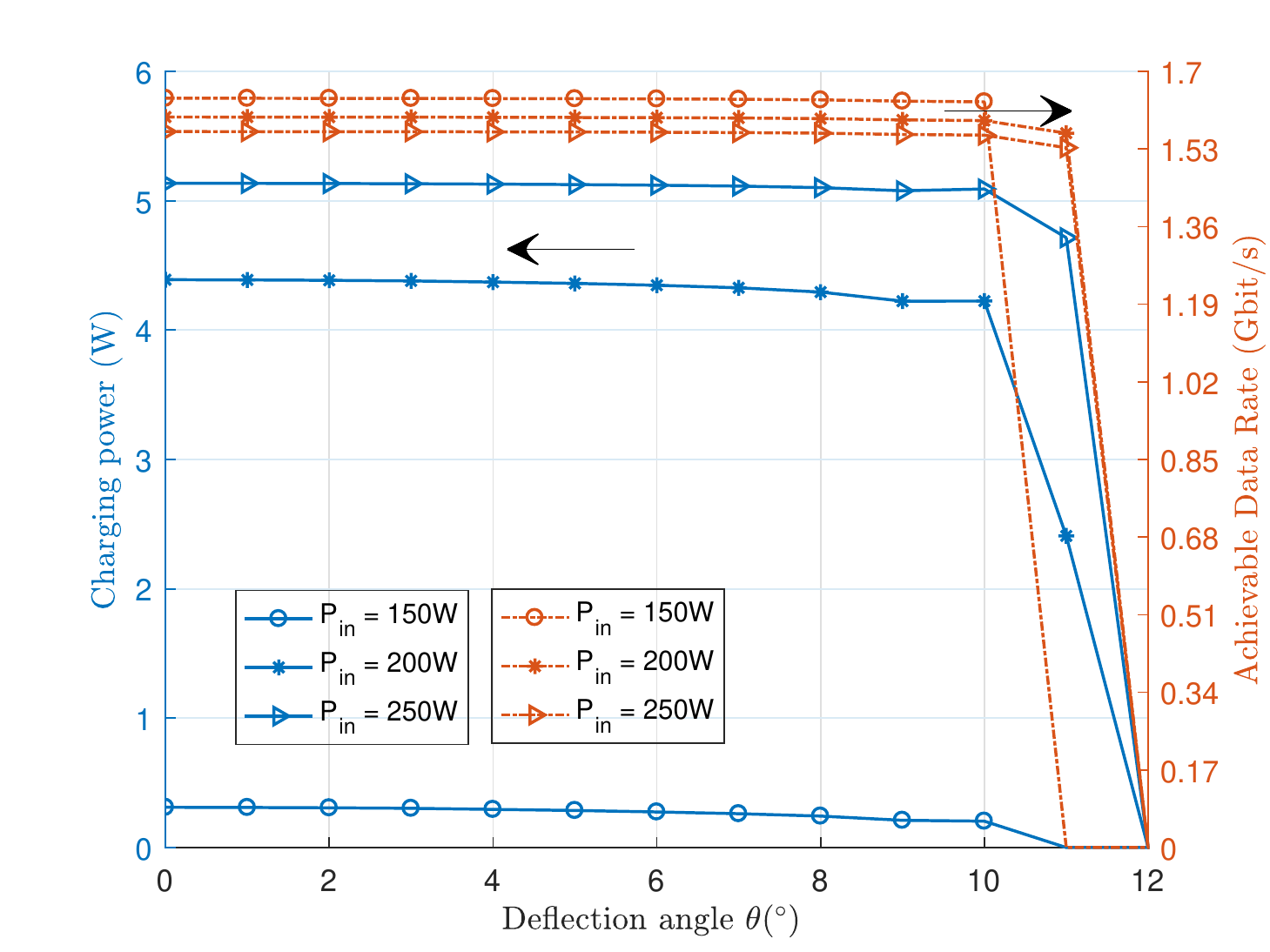}
    \caption{Charging power and achievable data rate as a function of deflection angle $\theta$ with different input power under transmission distance $L=3$m.}
    \label{f:Theta}
\end{figure}
In Fig.~\ref{f:Theta}, we analyze the relationship between the charging power/achievable data rate and the deflection angle $\theta$. The energy-data transfer performance shows the similar tendency with $\gamma$ as a function of $\theta$. The maximum operational deflection angle between the transceiver with $P_{\text{in}} = 200$W, $250$W is $12^{\circ}$. Moreover, as the input power increases, the charging power increases and the achievable data rate decreases, and if the input power is low, there will be no charging power output from the PV. Furthermore, the proposed RB-SLIPT system shows super performance of $5$W wireless charging power and $1.5$Gbit/s data rate with the capability of large moving range.

\section{Conclusions}
\label{sec:Conclusion}
In this paper, we designed an RB-SLIPT system which can deliver multi-Watt power to the mobile receivers with self-alignment characteristic. We at first established a mobile transmission channel model to reveal the mobility mechanism and quantitatively evaluate the energy distribution in the channel. Then we proposed an exemplary SLIPT design to transfer energy and data using PV and APD respectively. Finally, we analyze the impacts of moving factors on the energy/data transfer performance (charging power and achievable data rate). Numerical analysis illustrates that SMIPT is a practical solution for realizing mobile energy supply and data transfer.

Several interesting topics are worthy of further investigation in the future: 1) design and analysis for SWIPT with multiple receivers; 2) methods of enhancing deliverable power and data rate.

\bibliographystyle{IEEEtran}
\bibliographystyle{unsrt}
\bibliography{Reference}
\end{document}